\documentclass[preprint,showpacs,floatfix]{revtex4}
\usepackage[dvips]{graphicx}
\begin{document}
\title{$N_p N_n$ Scheme Based on New Empirical Formula \\for Excitation Energy}
\author{Jin-Hee \surname{Yoon}}
\author{Eunja \surname{Ha}}
\author{Dongwoo \surname{Cha}}
\email{dcha@inha.ac.kr}
\thanks{Fax: +82-32-866-2452}
\affiliation{Department of Physics, Inha University, Incheon
402-751, Korea}
\date{July 25, 2007}

\begin{abstract}
We examine the $N_p N_n$ scheme based on a recently proposed simple
empirical formula which is highly valid for the excitation energy
of the first excited natural parity even multipole states in
even-even nuclei. We demonstrate explicitly that the $N_p N_n$
scheme for the excitation energy emerges from the separate
exponential dependence of the excitation energy on the valence
nucleon numbers $N_p$ and $N_n$ together with the fact that only a
limited set of numbers is allowed for the $N_p$ and $N_n$ of the
existing nuclei.
\end{abstract}


\maketitle

\begin{figure}[b]
\centering
\includegraphics[width=12.0cm,angle=0]{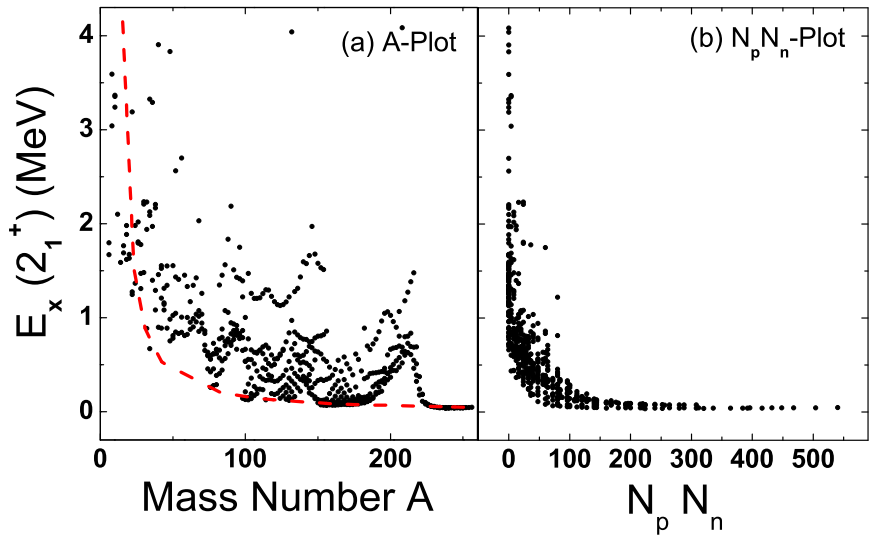}
\caption{A typical example demonstrating the $N_p N_n$ scheme. The
excitation energies of the first $2^+$ states in even-even nuclei
are plotted (a) against the mass number $A$ and (b) against the
product $N_p N_n$. The dashed curve in part (a) represents the
bottom contour line which is drawn by the first term $\alpha
A^{-\gamma}$ of Eq.\,(\ref{E}). The excitation energies are quoted
from Ref. 4.} \label{fig-1}
\end{figure}

The valence nucleon numbers $N_p$ and $N_n$ have been frequently
adopted in parameterizing various nuclear properties
phenomenologically over more than the past four decades. Hamamoto
was the first to point out that the square roots of the ratios of
the measured and the single particle $B(E2)$ values were
proportional to the product $N_p N_n$ \cite{Hamamoto}. It was
subsequently shown that a very simple pattern emerged whenever the
nuclear data concerning the lowest collective states was plotted
against $N_p N_n$ \cite{Casten1}. This phenomenon has been called
the $N_p N_n$ scheme in the literature \cite{Casten2}. For example,
when the measured excitation energies $E_x (2_1^+)$ of the first
excited $2^+$ states in even-even nuclei were plotted against the
mass number $A$ ($A$-plot), we got data points scattered irregularly
over the $E_x$-$A$ plane as seen in Fig.\,\ref{fig-1}(a). However,
we suddenly had a very neat rearrangement of the data points by just
plotting them against the product $N_p N_n$ ($N_pN_n$-plot) as shown
in Fig.\,\ref{fig-1}(b). A similar simplification was observed not
only from $E_x (2_1^+ )$ but also from the ratio $E_x (4_1^+)/ E_x
(2_1^+)$ \cite{Casten3,Casten4,Casten5}, the transition probability
$B(E2; 2_1^+ \rightarrow 0^+)$ \cite{Casten6}, and the quadrupole
deformation parameter $e_2$ \cite{Casten7}.

The chief attraction of the $N_PN_n$ scheme is twofold. One is the
fact that the simplification in the graph occurs marvelously every
time the $N_pN_n$ plot is drawn. The other attraction is the
universality of the pattern, namely the exactly same sort of graphs
appears even at different mass regions \cite{Casten1}. Since the
performance of the $N_pN_n$ scheme has been so impressive, many
expected that the residual valence proton-neutron (p-n) interaction
must have been the dominant controlling factor in the development of
collectivity in nuclei and that the product $N_pN_n$ may represent
an empirical measure of the integrated valence p-n interaction
strength \cite{Casten2}. Also, the importance of the p-n interaction in determining the structure of nuclei has long been pointed out by many authors \cite{deShalit,Talmi,Heyde,Casten8,Zhang,Federmann,Dobaczewski}.

\begin{figure}[t]
\centering
\includegraphics[width=12.0cm,angle=0]{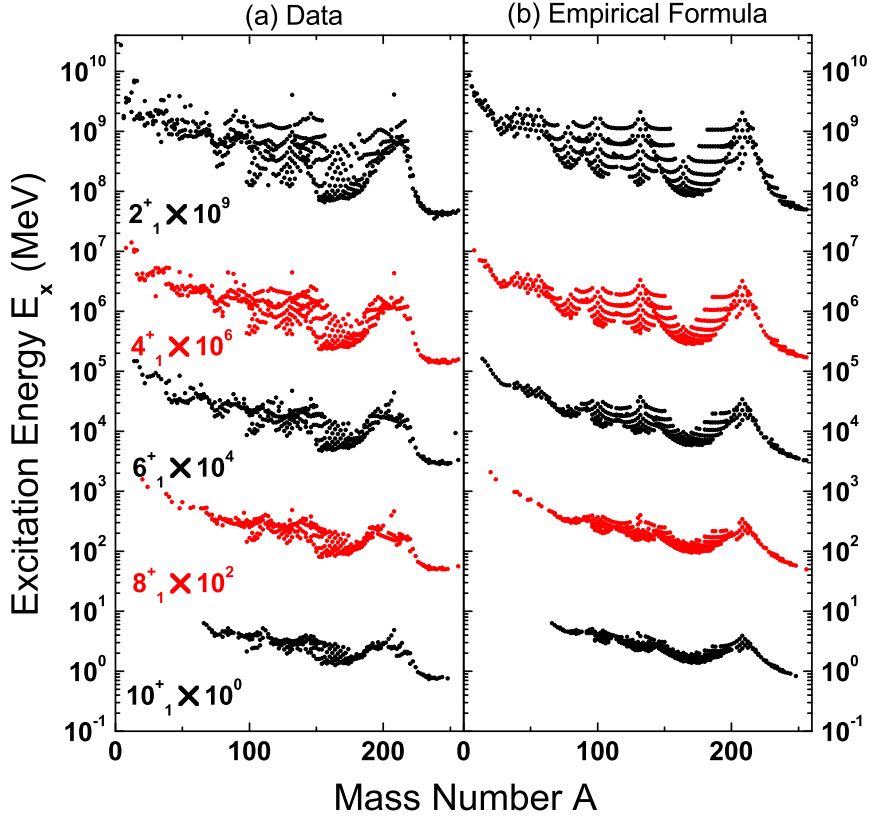}
\caption{Excitation energies of the first excited natural parity
even multipole states. Part (a) shows the measured excitation
energies while part (b) shows those calculated by the empirical
formula given by Eq.\,(\ref{E}). The measured excitation energies
are quoted from the compilation in Raman {\it et al}. for $2_1^+$
states \cite{Raman} and extracted from the Table of Isotopes,
8th-edition by Firestone {\it et al}. for other multipole states
\cite{Firestone}.} \label{fig-2}
\end{figure}

In the meantime, we have recently proposed a simple empirical
formula which describes the essential trends of the excitation
energies $E_x (2_1^+)$ in even-even nuclei throughout the periodic
table \cite{Ha1}. This formula, which depends on the valence nucleon
numbers, $N_p$ and $N_n$, and the mass number $A$, can be expressed
as
\begin{equation} \label{E}
E_x = \alpha A^{-\gamma} + \beta \left[ \exp ( - \lambda N_p ) +
\exp ( - \lambda N_n ) \right]
\end{equation}
where the parameters $\alpha$, $\beta$, $\gamma$, and $\lambda$ are
fitted from the data. We have also shown that the source, which
governs the $2_1^+$ excitation energy dependence given by
Eq.\,(\ref{E}) on the valence nucleon numbers, is the effective
particle number participating in the residual interaction from the
Fermi level \cite{Ha2}. Furthermore, the same empirical formula can
be applied quite successfully to the excitation energies of the
lowest natural parity even multipole states such as $4_1^+$,
$6_1^+$, $8_1^+$, and $10_1^+$ \cite{Kim}. It can be confirmed by
Fig.\,\ref{fig-2} where the measured excitation energies in part (a)
are compared with those in part (b) which are calculated by
Eq.\,(\ref{E}). The values of the parameters adopted for
Fig.\,\ref{fig-2}(b) are listed in Table \ref{tab-1}.

\begin{figure}[b]
\centering
\includegraphics[width=12.0cm,angle=0]{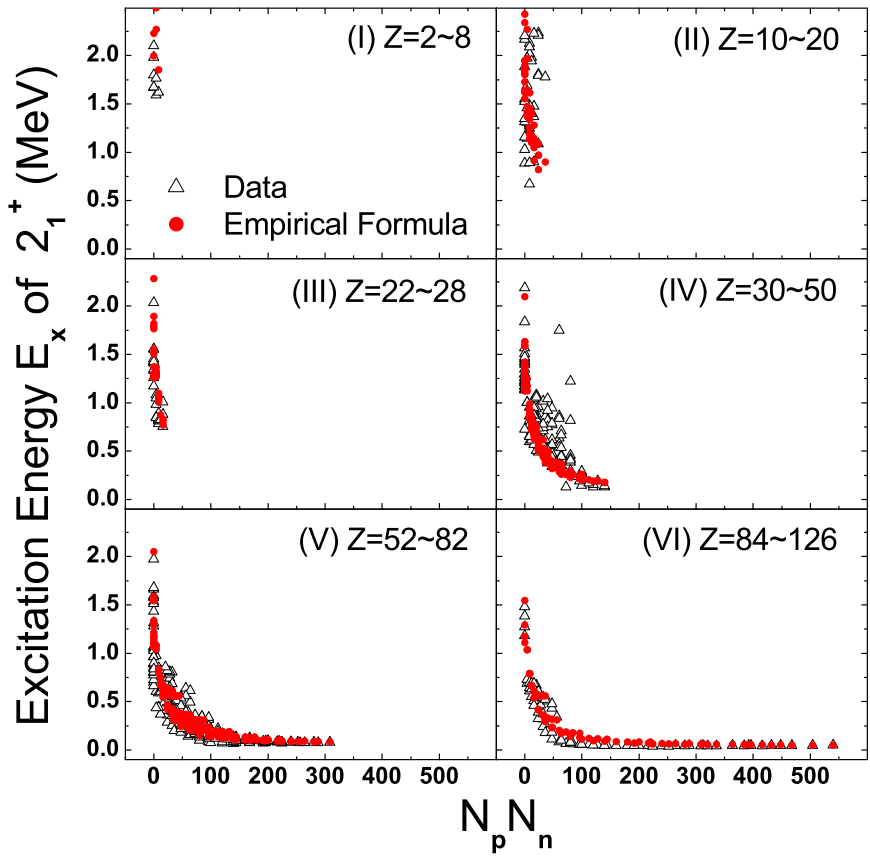}
\caption{The $N_p N_n$-plot for the excitation energies of the first
$2^+$ states using both the data (open triangles) and the empirical
formula (solid circles). The plot is divided into six panels each of
which contains plotted points that come from each one of the proton
major shells.} \label{fig-3}
\end{figure}

In this study, we want to further elucidate about our examination of
the $N_p N_n$ scheme based on the empirical formula, Eq.\,(\ref{E}),
for $E_x (2_1^+)$. Our goal is to clarify why $E_x (2_1^+)$ complies
with the $N_p N_n$ scheme although the empirical formula, which
reproduces the data quite well, does not depend explicitly on the
product $N_p N_n$.

\begin{table}[t]
\begin{center}
\caption{Values adopted for the four parameters in Eq.\,(\ref{E})
for the excitation energies of the following multipole states:
$2_1^+$, $4_1^+$, $6_1^+$, $8_1^+$, and $10_1^+$.}
\begin{tabular}{ccccc}
\hline\hline
Multipole&~~~$\alpha$(MeV)~~~&~~~$\beta$(MeV)~~~&~~~~~~$\gamma$~~~~~~&~~~~~~$\lambda$~~~~~~\\
\hline
$2_1^+$&34.9&1.00&1.19&0.36\\
$4_1^+$&94.9&1.49&1.15&0.30\\
$6_1^+$&441.4&1.51&1.31&0.25\\
$8_1^+$&1511.5&1.41&1.46&0.19\\
$10_1^+$&2489.0&1.50&1.49&0.17\\
\hline\hline
\end{tabular}
\label{tab-1}
\end{center}
\end{table}

\begin{table}[b]
\begin{center}
\caption{The maximum value of $N_p N_n$ and the minimum value of
$E_x$ for each major shell in Fig.\,\ref{fig-3} are indicated here.
The numbers in the parenthesis represent $E_x$ calculated by the
empirical formula given by Eq.\,(\ref{E}).}
\begin{tabular}{cccc}
\hline\hline
Major Shell~~~~&~~~$Z$~~~&~~Max. $N_p N_n$~~&~Min. $E_x$ (MeV)~\\
\hline
I&$2 \sim 8$&$8$&$1.59\,(1.85)$\\
II&$10 \sim 20$&$36$&$0.67\,(0.82)$\\
III&$22 \sim 28$&$16$&$0.75\,(0.77)$\\
IV&$30 \sim 50$&$140$&$0.13\,(0.18)$\\
V&$52 \sim 82$&$308$&$0.07\,(0.08)$\\
VI&$84 \sim 126$&$540$&$0.04\,(0.05)$\\
\hline\hline
\end{tabular}
\label{tab-2}
\end{center}
\end{table}

First, we check how well the empirical formula does meet the
requirements of the $N_p N_n$ scheme. In Fig.\,\ref{fig-3}, we
display the $N_p N_n$-plot for the excitation energies of the first
$2^+$ states using both the data (empty triangles) and the empirical
formula (solid circles). We show them with six panels. Each panel
contains plotted points from nuclei which make up the following six
different proton major shells: (I) $2 \le Z \le 8$, (II) $10 \le Z
\le 20$, (III) $22 \le Z \le 28$, (IV) $30 \le Z \le 50$, (V) $52
\le Z \le 82$, and (VI) $84 \le Z \le 126$. From this figure, we can
see an intrinsic feature of the $N_p N_n$-plot, namely, the plotted
points have their own typical location in the $E_x$-$N_p N_n$ plane
according to which major shell they belong. For example, the plotted
points of the first three major shells I, II, and III occupy the far
left side part of the $E_x$-$N_p N_n$ plane in Fig.\,\ref{fig-3}
since their value of the product $N_p N_n$ does not exceed several
tens. On the contrary, the plotted points of the last major shell VI
extend to the far right part of the $E_x$-$N_p N_n$ plane along the
lowest portion in Fig.\,\ref{fig-3}. This is true since their value
of the excitation energy $E_x$ is very small and also their value of
$N_p N_n$ reaches more than five hundreds. We present specific
information such as the maximum value of $N_p N_n$ and the minimum
value of $E_x$ in Table\,\ref{tab-2} for the plotted points which
belong to each major shell in Fig.\,\ref{fig-3}. There are two
numbers for each major shell in the last column of
Table\,\ref{tab-2} where one number is determined from the data and
the other number in parenthesis is calculated by the empirical
formula. We can find that those two numbers agree reasonably well.
We also find in Fig.\,\ref{fig-3} that the results, calculated by
the empirical formula (solid circles), meet the requirement of the
$N_p N_n$ scheme very well and agree with the data (empty
triangles) satisfactorily for each and every panel.

\begin{figure}[t]
\centering
\includegraphics[width=12.0cm,angle=0]{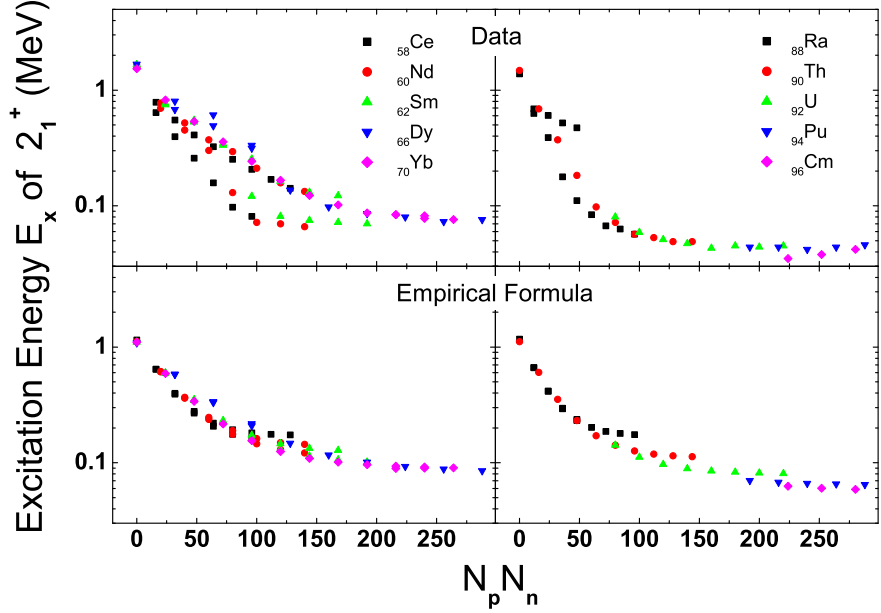}
\caption{Extract from Fig.\,\ref{fig-3} for some typical nuclei which belong to the rare earth elements. Different symbols are used to denote excitation energies of individual nuclei.} \label{fig-3a}
\end{figure}

In order to make more detailed comparison between the measured and calculated excitation energies, we expand the largest two major shells V and VI of Fig.\,\ref{fig-3} and redraw them in Fig.\,\ref{fig-3a} for some typical nuclei which belong to the rare earth elements. The upper part of Fig.\,\ref{fig-3a} shows the data and the lower part of the same figure exhibits the corresponding calculated excitation energies. We can confirm that the agreement between them is reasonable even though the calculated excitation energies somewhat overestimate the data and also the empirical formula can not separate enough to distinguish the excitation energies of the two isotopes with the same value of the product $N_pN_n$ for some nuclei.

\begin{figure}[t]
\centering
\includegraphics[width=12.0cm,angle=0]{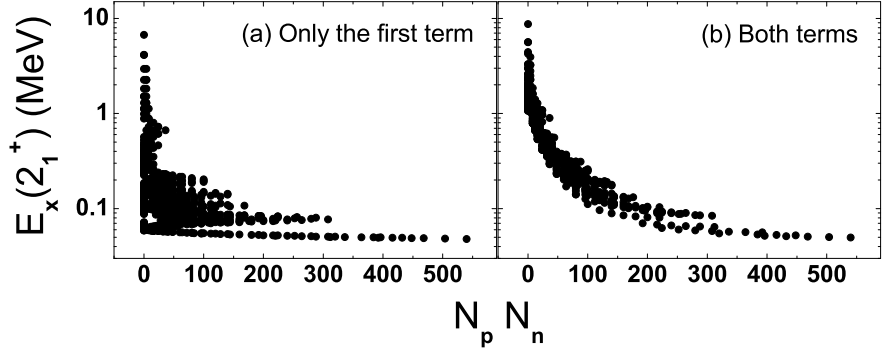}%
\caption{The $N_pN_n$-plot of the calculated first excitation energy $E_x$ of $2^+$ states. The excitation energies $E_x$ are calculated by (a) using only the first term and (b) using both terms of Eq.\,(\ref{E}).}
\label{fig-4}
\end{figure}

According to the empirical formula given by Eq.\,(\ref{E}), the
excitation energy $E_x$ is determined by two components: one is the
first term $\alpha A^{-\gamma}$ which depends only on the mass
number $A$ and the other is the second term $\beta [ \exp (- \lambda
N_p ) + \exp (- \lambda N_n )]$ which depends only on the valence
nucleon numbers, $N_p$ and $N_n$. Let us first draw the
$N_pN_n$-plot of $E_x (2_1^+)$ by using only the first term $\alpha
A^{-\gamma}$. The results are shown in Fig.\,\ref{fig-4}(a) where we
can find that the plotted points fill the lower left corner of the
$E_x$-$N_pN_n$ plane leaving almost no empty spots. These results
simply reflect the fact that a large number of nuclei with different
mass numbers, values of $A$, can have the same value of $N_pN_n$.
Now we draw the same $N_pN_n$-plot by using both of the two terms in
Eq.\,(\ref{E}). We display the plot of the calculated excitation energies in Fig.\,\ref{fig-4}(b) which is just the same sort of graph of the measured excitation energies shown in Fig.\,\ref{fig-1}(b) except that the type of scale for $E_x$ is changed from linear to log. By comparing Fig.\,\ref{fig-4} (a) and (b), we find that the second term of Eq.\,(\ref{E}), which depends on the valence nucleon numbers, $N_p$ and $N_n$, pushes the plotted points up in the direction of higher excitation energies and arranges them to comply with the $N_pN_n$ scheme.

It is worthwhile to note the difference between the $A$-plot and the
$N_pN_n$-plot. The graph drawn by using only the first term of
Eq.\,(\ref{E}) becomes a single curve in the $A$-plot as shown in
Fig.\,\ref{fig-1}(a) with the dashed curve. It becomes scattered
plotted points in the $N_pN_n$-plot as can be seen from
Fig.\,\ref{fig-4}(a). Now, by adding the second term of
Eq.\,(\ref{E}) in the $A$-plot, the plotted points are dispersed as
shown in the top graph of Fig.\,\ref{fig-2}(b) which corresponds to the measured data points in Fig.\,\ref{fig-1}(a); while by adding the same second term in $N_pN_n$-plot, we find a very neat rearrangement of the plotted points as shown in Fig.\,\ref{fig-4}(b). Thus, the same second term plays the role of spreading plotted points in the $A$-plot while it plays the role of collecting them in the $N_pN_n$-plot.

\begin{figure}[b]
\centering
\includegraphics[width=12.0cm,angle=0]{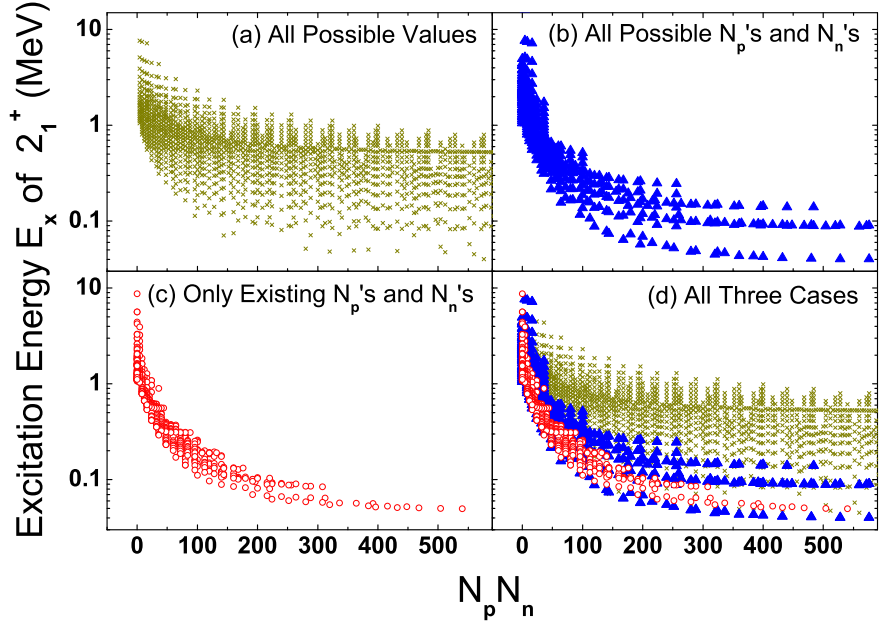}%
\caption{The $N_pN_n$-plot of the first excitation energy of the
$2^+$ states calculated by the empirical formula given by
Eq.\,(\ref{E}) using the following three different conditions on the
exponent $N_p$ and $N_n$: (a) $N_p$ and $N_n$ can have any even
numbers as long as they satisfy $N_p+N_n \le A$. (b) $N_p$ and $N_n$
can have any number that is allowed for the valence nucleon numbers.
(c) $N_p$ and $N_n$ can have numbers which are allowed for the
actually existing nuclei. (d) All of the previous three cases are
shown together.} \label{fig-5}
\end{figure}

However, this mechanism of the second term alone is not sufficient
to explain why the empirical formula given by Eq.\,(\ref{E}) which
obviously does not depend on $N_pN_n$ at all, can show the
characteristic feature of the $N_pN_n$ scheme. In order to shed
light on this question, we calculate the excitation energy
$E_x(2_1^+)$ by the following three different conditions on the
exponents, $N_p$ and $N_n$, of the second term in Eq.\,(\ref{E}).
First, let $N_p$ and $N_n$ have any even numbers as long as they
satisfy $N_p + N_n \le A$. The resulting excitation energy $E_x$ is
plotted against $N_pN_n$ in Fig.\,\ref{fig-5}(a). Next, let $N_p$
and $N_n$ have any numbers that are allowed for the valence nucleon
numbers. For example, suppose the three numbers of a plotted point are $A=90$, $N_p=40$, and $N_n=50$ in the previous case. For the fourth major shell IV in Table\,\ref{tab-2}, the valence proton number for the nucleus with the atomic number $Z=40$ is 10 and the valence neutron number for the nucleus with the neutron number $N=50$ is 0. Therefore, we assign $N_p=10$ and $N_n=0$ instead of 40 and 50, respectively. The excitation energy $E_x$, calculated under such a
condition, is plotted against $N_pN_n$ in Fig.\,\ref{fig-5}(b).
Last, we take only those excitation energies which are actually
measured among the excitation energies shown in
Fig.\,\ref{fig-5}(b). The results are shown in Fig.\,\ref{fig-5}(c),
which is, of course, exactly the same as shown in
Fig.\,\ref{fig-4}(b). From Fig.\,\ref{fig-5}(d) where all the three
previous plots (a), (b), and (c) are placed together, we can observe
how the $N_pN_n$ scheme emerges from the empirical formula given by
Eq.\,(\ref{E}) even though this equation does not depend on the
product $N_pN_n$ at all. On one hand, the two exponential terms
which depend on $N_p$ and $N_n$ separately push the excitation
energy $E_x$ upward as discussed with respect to Fig.\,\ref{fig-4}.
On the other hand, the restriction on the values of the valence
nucleon numbers $N_p$ and $N_n$ of the actually existing nuclei
determines the upper bound of the excitation energy $E_x$ as
discussed regarding Fig.\,\ref{fig-5}.

\begin{figure}[t]
\centering
\includegraphics[width=12.0cm,angle=0]{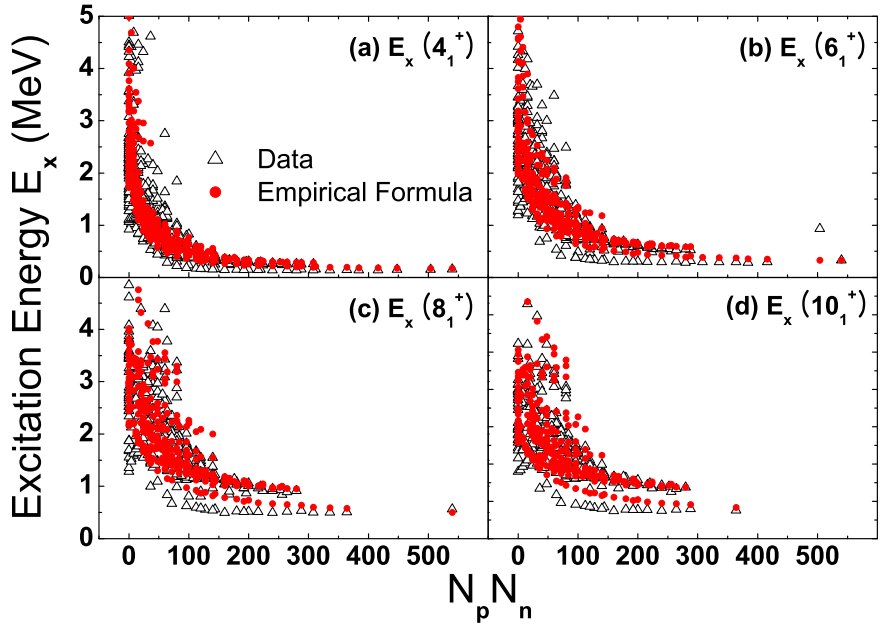}%
\caption{The $N_pN_n$-plot for the first excitation energies of the
natural parity even multipole states (a) $4_1^+$, (b) $6_1^+$, (c)
$8_1^+$, and (d) $10_1^+$ using both the measured data (open
triangles) and the empirical formula (solid circles). These graphs
are just the $N_pN_n$-plot versions of the $A$-plot shown in
Fig.\,\ref{fig-2} with exactly the same set of data points.}
\label{fig-6}
\end{figure}

Finally, we show the $N_pN_n$-plots of the first excitation energies
for (a) $4_1^+$, (b) $6_1^+$, (c) $8_1^+$, and (d) $10_1^+$ states
in Fig.\,\ref{fig-6}. The measured excitation energies are
represented by the empty triangles and the calculated ones from the
empirical formula, Eq.\,(\ref{E}), are denoted by solid circles.
These graphs are just the $N_pN_n$-plot versions of the $A$-plot
shown in Fig.\,\ref{fig-2} with exactly the same set of plotted points. We can learn from Fig.\,\ref{fig-6} that the same kind of $N_pN_n$ scheme observed in the excitation energy of $2_1^+$ states is also functioning in the excitation energies of other natural parity even multipole states. We can also find from Fig.\,\ref{fig-6} that the calculated results, using the empirical formula, agree with the measured data quite well. Moreover, it is interesting to find
from Fig.\,\ref{fig-6} that the width in the central part of the
$N_pN_n$-plot is enlarged as the multipole of the state is
increased. The origin of this enlargement in the empirical formula
can be traced to the parameter $\alpha$ of the first term in
Eq.\,(\ref{E}). The value of $\alpha$ is monotonously increased from
$34.9\,{\rm MeV}$ for $E_x(2_1^+)$ to $2489.0\,{\rm MeV}$ for
$E_x(10_1^+)$ as can be seen in Table\,\ref{tab-1}.

In summary, we have examined how the recently proposed empirical
formula, Eq.\,(\ref{E}), for the excitation energy $E_x(2_1^+)$ of
the first $2_1^+$ state meets the requirement of the $N_pN_n$ scheme
even though it does not depend on the product $N_pN_n$ at all. We have demonstrated explicitly that the structure of the empirical formula itself together with the restriction on the values of the valence nucleon numbers $N_p$ and $N_n$ of the actually existing nuclei make the characteristic feature of the $N_pN_n$ scheme appear. Furthermore, our result shows that the composition of the
empirical formula, Eq.\,({\ref{E}), is in fact ideal for revealing
the $N_pN_n$ scheme. Therefore it is better to regard the $N_pN_n$
scheme as a strong signature suggesting that this empirical formula
is indeed the right one. As a matter of fact, this study
about the $N_pN_n$ scheme has incidentally exposed the significance
of the empirical formula given by Eq.\,(\ref{E}) as a universal
expression for the lowest collective excitation energy. A more
detailed account of the empirical formula for the first excitation
energy of the natural parity even multipole states in even-even
nuclei will be published elsewhere \cite{Kim}. However, it has been well established that the $N_pN_n$ scheme holds not only for the lowest excitation energies $E_x (2_1^+)$ but also for the transition strength $B(E2)$ \cite{Casten6}. Unfortunately, our empirical study intended to express only the excitation energies in terms of the valence nucleon numbers. The extension of our study to include the $B(E2)$ values in our parametrization is in progress.

\begin{acknowledgments}
This work was supported by an Inha University research grant.
\end{acknowledgments}

\end{document}